\documentstyle{article}

\begin{document}

{\bf HYPERGRAVITY AND CATEGORICAL FEYNMANOLOGY }

\bigskip

by Louis Crane

Mathematics department, KSU.

\bigskip

{\bf Abstract:} We propose a new line of attack to create a finite quantum theory which includes general
relativity and (possibly) the standard model in its low energy
limit. The theory would emerge naturally from the categorical
approach. The traces of morphisms from the category of representations
we use to construct the state sum also admit interpretation as Feynman diagrams, so the
non-categorically minded physicist may think of the models as 
combinatorial expressions in Feynman integrals, reflecting the
topology of the triangulated manifold. The Feynman picture of the
vacuum would appear as a low energy limit of the theory. 

The fundamental dynamics of the
theory are determined by a Topological Quantum Field
Theory expanded around a conjectured geometric quasi-vacuum. 

Although
the model is a 4 dimensional state sum, it
would have a stringlike ten dimensional perturbation theory.

The motivation for the
name ``hypergravity'' is the existence of an infinite tower of
alternating fermionic and bosonic partners of the gravitational field
in the theory, with an n=2 chiral supersymmetry functor connecting
them. It is remarkable that a supersymmetry appears in the high energy
sector of a theory not founded on supergeometry.

\bigskip

{\bf 1. Introduction}

\bigskip

The purpose of this paper is to begin a new attempt to produce
a quantum theory of nature, including both general relativity and
particle physics, out of the categorical approach. [CKY1,BC1,BC2].

The categorical approach is a method to produce a type of discrete theory, rather like a
lattice QFT, but put on a triangulated manifold, instead of a lattice,
in order to make it applicable to curved spacetime.

One replaces the
fields and Lagrangian of a QFT by an abstract algebraic structure and its
operations.  A basis for the algebraic structure is used to label sites
on the triangulation, and the operations are used to combine these to
make numbers called ``local contributions''.These are then combined into a state
sum, which can be viewed as closely analogous to a path integral.
The analogy is that the elements of the structure are like local
fields, while the structure itself is a sort of Lagrangian density.

In higher dimension than 2, the algebraic structure is
often some sort of tensor category and the structure which is used
to create the QFT is its associator or something similar.

The local contributions from the relevant tensor category are in fact the
evaluations of Feynman diagrams for a special QFT on hyperbolic
space[F-K,BC2]. As we shall argue below, this suggests that the Feynmanology of
the matter fields in the model we are proposing may appear directly as
traces of morphisms in the category of representations from which we
construct the theory.

The current paper is
not self contained. It draws heavily on the categorical state sum
models for 
$B \wedge F$
theory in [CKY1], and the Euclidean and Lorentzian signature state sum
models in [BC1,BC2]. We assume that the reader is familiar with those
constructions, as well as the fundamentals of category theory. The
entire motivation for the proposal is specific features of these
models, and of the tensor category from which they are constructed.

The construction of a four
dimensional topological state sum (TSS) in [CKY1] assumes a complete tensor
category, i.e. one which is closed under the tensor product. In [BC1,2],
models were proposed for four dimensional discrete quantum general relativity in Euclidean
and Lorentzian signatures respectively. Both models were constrained
versions of models of the topological type in [CKY1], which were
constructed by restricting to a suitable subcategory which was not a
tensor subcategory. The Euclidean model [BC1] was obtained by
restricting to the irreducible representations of $U_q so(4)$ which
had equal half integer indices for the left and right actions of $U_q
so(3)$ , which we call ``balanced''. The Lorentzian model was a
similar restriction of the category of representations of the Quantum
Lorentz Algebra [BR 1,2], specifically to the irreducible representations
of the form $R_{0, p}$. (the general irreducible representation of the
QLA is of the form $R_{k,p}$ for k a half integer.) Since these models
are not based on closed tensor categories, they are not topological,
i.e. they depend on the triangulation used.

In both cases the construction is motivated by the idea that the
constraint corresponds to the quantization of the condition that the
bivectors on the 2-simplices of a triangulated 4-manifold are simple.

This leaves two difficult problems in the way of developing a serious
physical theory: 

\bigskip

{ \bf 1. finding the limiting behavior as the triangulation is refined,
i.e. the continuum limit }

\bigskip

and

\bigskip

{\bf 2. adding the matter fields.}

\bigskip

The behavior of the topological model as the triangulation is refined
is not problematical at all. The physical quantities are triangulation
independent, so we are already at the continuum limit. The price we
pay for this is that there are no local degrees of freedom, not even
geometrical ones. The categorical constraint has the effect of
breaking topological invariance, so the ``smaller'' state sum has a
``larger'' state space.

Several authors have tried to make a continuum limit for the models of
[BC1,2] or closely related ones by summing over triangulations or
slightly more general combinatorial situations [R-R1,Ba]. The convergence
problem for such an approach seems formidable, however.

Our hope is that the two problems have a single solution. 

What we want to propose is that the fundamental dynamics of our
universe is actually determined by a categorical topological state sum, but that we
are in a sort of ``bubble'', i.e. a region in which labels in the sum
are restricted to a subspace of the full labelling set
of objects of the category which has a relative stability under the
dynamics induced by the TSS. This is similar to the suggestion of
Witten [W] that the fundamental theory has a ``topological phase'' but
we fluctuated out of it, except that it is rather a TQFT that has a
nontopological quasiphase.

The constrained subtheory would be
a summation over some of the labels for the whole theory, but subject to a
constraint which has the property that if we begin a state sum with
initial conditions satisfying the constraint, the dynamics of the
larger TSS would only produce a very small contribution of labels
outside the constraint for a long
time as measured by a clock constructed within the constrained label
space, provided the curvature of the geometry defined by the labelling
within the constrained subsystem was small compared to the Planck
scale.

Let us formalize matters:

{\bf ``DEFINITION'':} A {\bf geometric quasi-vacuum} is a subspace S of
the total labelling space T for a TSS such that
\bigskip

1. The constrained labellings can be interpreted as defining a
discrete geometry on the underlying manifold.

\bigskip

2. If we calculate the time evolution using the entire TSS, but on
initial conditions inside the subspace S, the deviation from the
result of the constrained theory will be small provided the curvature
of the initial conditions is small in Planck units, for a long time as
measured by a clock defined in terms of labels from the set S.

\bigskip

{\bf CONJECTURE 1:} The Euclidean signature model of [BC1] is a geometric
quasi-vacuum for the TSS of [CKY1] with category Rep $U_q so(4)$ , i.e. the
space of balanced labels is a suitable S in the above definition. The
Lorentzian model in [BC2], but with q taken to a root of unity, is a
geometrical quasi vacuum of a TSS associated to a version of the QLA
with q also a root of unity.

\bigskip

{\bf CONJECTURE 2:} The set of geometric quasi-vacua associated to
the TSSs of conjecture 1 is actually very large, including examples
related to string vacua.

\bigskip

Motivation for these conjectures will appear in the bulk of the paper
below.

\bigskip

What we are proposing is that the matter fields we see in the universe
are the low energy limit of the other labels in the TSS, besides those
that satisfy the constraints for the model for general relativity. If we are in a ``bubble'',
i.e. a region where the labels satisfying the constraints
predominate, small fluctuations of the extra labels would appear to
propagate through the geometrical background defined by the included
labels.

An observer in a region of a TSS which fluctuated into a geometric
quasi vacuum would seem to see the world obey the laws of the
constrained theory. In particular if the constrained theory
approximated the laws of GR coupled to suitable matter, it could
easily settle into an expanding solution, so a Planck scale initial
fluctuation could appear to become vastly larger to an internal
observer. If such an observer attempted to study a region
where spacetime was highly curved, then the labels excluded from the
constraint would begin to appear, adding at first to the field content
of the observed theory. Paradoxically, however, if the observer
entered a region which strayed too far from the curvature limitations,
and hence from the constrained theory, the additional hierarchy of
``fields'' would produce a regime in which no local excitations could
exist at all anymore, i.e. a topological theory. One way to understand
how this would happen is that the new sets of labels which would
appear are linked by a supersymmetry functor, so that the corrections
which would appear if one refined the triangulation would cancel out
in a telescoping supersymmetry.

The way I envision this is that what would appear would be a thermal bath of
an ever increasing variety of particles, as predicted by the
semiclassical theory in regions of curvature. Eventually,this would reach a critical
temperature and make all measurement impossible, destroying any
observer in the region.

Put differently, there would not exist a true continuum quantum theory
of pure gravity, because at sufficiently high energy densities matter
would be formed around the curvature loci. Matter plus gravity would
not have a continuum quantum theory, because at still higher
temperature the matter fields and metric fields would merge into a
topological state sum. Since the higher partners of the labels which
give gravity are related by a supersymmetry functor, their
contributions to the state sum would telescopically cancel as we
refined the triangulation into the region where they became
important. Thus, an effect analogous to the no renormalization theorem
for supersymmetric theories would cause the state sum to become
topological.

This is rather similar to the idea of SUSY breaking, except in a
discrete picture, rather than for continuum fields. It is remarkable
that the category of unitary representations of the ordinary Lorentz
group has a decomposition into partners of the subcategory associated to
GR connected by a supersymmetry functor. Infinite supersymmetric hierarchies
do not seem to appear in continuum field theories.

The region of spacetime before the fluctuation happened would seem to
have a completely degenerate geometry and a thermal state at the
Planck temperature to an observer inside the bubble. This would be a
plausible scenario for a big bang.

Any region inside the bubble where the curvature became high enough
would also begin to see labels from outside the constraint space,
which would appear similar to Hawking radiation, until the Planck
temperature was reached.

A region
within the bubble which appeared so small that any subdivision of it
would be dominated by labellings with large curvature would
effectively leave the bubble. An attempt by an observer to probe it
would yield a thermal state, and no finer determinateness as to the
geometry of the small region. Thus the only continuum limit for the
theory would be the topological theory.

The ``definition'' of a geometric quasi-vacuum given above is
motivated by a physical picture, and will need considerable refinement
to make it mathematically precise enough for further analysis. We will
try to argue that various specific facts about the models of [CKY1 BC1,2]
make this picture plausible. {\bf It is with quasi-vacua of the specific
TSS related to the unitary representations of the QLA and $U_qso(4)$ that we shall be concerned.}
For brevity, and for reasons discussed below we shall refer to them as
``hypergravity''.

In order to advance the program of this paper, the first necessary
step will be to make a precise formulation of the definition of
quasi-vacuum in terms of the actual categorical state sum. A natural
procedure would be to constrain the labels along a 3D hypersurface in
the triangulated 4-manifold to lie within the theory, and to constrain
the labels on the 2-simplices incident to the hypersurface both to lie
within the constraint subcategory and to correspond geometrically to a
small time. One could investigate the behaviour of the state sum for
some relatively simple constraint subcategory, then try to prove a
quantitative form of the conjecture. Lastly, one could try to
generalize the behaviour to a large family of constraint
subcategories. We are then conjecturing that a very large family of
subspaces of the set of labels would satisfy such a stability
condition. If our intuition is correct, we would end up with a
picture of the vacuum of the TQFT as a froth of many different types
of bubble.

What we want to do in this paper is to make a preliminary analysis of
the prospects of this direction of development. There are two
points which seem to us to make it worthy of study.

\bigskip

1. There seems to be a natural picture in which the low energy fields
produced by the excluded labels would reproduce the standard
model. This is because the algebraic structure of the category we are
proposing to use makes very natural contact with the ideas of the
Connes-Chamseddine model [CC, Coq].

\bigskip

2. A perturbative treatment of the hypergravity model  
yields a picture of strings with either bosonic or fermionic labels
moving through a 10 dimensional curved space. It is not clear yet if these
are related to superstrings as we know them. Nevertheless, this is the
motivation for conjecture 2.

In the rest of this paper we will have to treat specific aspects of
the CSS's mentioned above, in order to see to what extent the
conjectures are justified.

\bigskip

{\bf 2. Representation theory of the QLA and the Hypergravity
Multiplet. Constraints and Quasi-Vacua.}

\bigskip

Let us begin by reviewing the basic facts concerning the
unitary representations of the classical Lie algebra so(3,1) which is
isomorphic to $sl(2,C)_R$. These were first studied by Gelfand, and
his collaborator Naimark [GN,N]. There is a fundamental family of
these, called the principal series. They can be realized as weighted
actions of the group of linear fractional transformations on the
Hilbert space $L^2(C)$. They form a two parameter family $R_{k,p}$,
where k is a half integer and p is a real number. For any three
representations $R_{k_i,p_i}$, i=1,2,3; there is a unique intertwining
operator $R_{k_1,p_1} \otimes R_{k_2,p_2} \rightarrow R_{k_3,p_3}$ up to
scalar multiple if $k_1+k_2+k_3$ is an integer, and there is no
intertwiner at all otherwise. The intertwiners are given explicitly in
[N].

The constraint for the model of [BC2] is that k=0; i.e. we consider
only a state integral over the 10J symbols for the $R_{0,p}$'s.

In order to get a finite model in [BC2], we passed to representations of
the quantum group associated to the Lorentz group, the QLA of [P-Wo, Pu, BR 1,2].
The representation theory of the QLA, for a real deformation parameter
q is well analyzed in the above papers, and is quite similar to the
picture for the ordinary Lorentz algebra, except that the continuous
parameter p only takes values in the interval [0,$ \frac {4 \pi}
{ln(q)} $]. The picture for intertwiners is also analogous to the result
cited for the Lorentz algebra. The result was that it was possible to
write down a state integral model using the representations of the
QLA, with all the integrals over finite measures, and hence finite.

Now we would like to think of the constrained model of [BC2] as a
submodel of a TSS. This is not quite possible for real values of q,
since the sum over the discrete parameter k would still be infinite.

However, it is natural to conjecture that if we passed to q a root of
unity, then there would be a truncation of the family of
representations $R_{k,p}$ at a sufficiently large k for q a fixed root
of unity. The reason to believe that is that the representation
$R_{k,p}$ decomposes as an ascending chain of representations of
$U_qsu(2)$ with the lowest spin in the chain given by k. Thus, since
the representations of $U_qsu(2)$ with spin greater than (lnq/4i)-1 have
quantum dimension 0 and decouple, so should the representations of the
QLA at the same k. Assuming this conjecture is correct, and it is
certainly not too hard to investigate, We can embed the Lorentzian
model of [BC2] as a constrained subsum of a TSS. We already know that
the Euclidean signature model of [BC1] can be embedded as a
subsum of a TSS, namely the model of [CKY1] with category $Rep U_q
so(4)$ for q a root of unity.

In either case, it is interesting to think how the rest of the
labelling category appears in relationship to the subcategory we are
using as a discrete model for gravity. We shall discuss the Lorentzian
signature, the Euclidean situation is very similar.
The labelling objects appear as a series of copies of the labels for
gravity, one partner set {  $R_{k,p}$ } for each half integer k. The couplings
for each partner set would be similar to the ones in the gravity
labelling set, but different sets would be coupled more weakly to one
another by some exponential factor involving the difference in k [BR
2].

Thus we would find a hierarchy of ``partners'' to gravity, alternately
bosonic and fermionic.

There is also a natural operation in the category which connects the
adjacent partner sets. Tensoring with a finite dimensional
representation (necessarily nonunitary) then projecting back to the
unitary category would send each $R_{k,p}$ to a combination of other
k's determined by the usual Clebsch Gordon formula, but only with the
identical p [BR3]. In particular, tensoring with either the (1/2,0) or
(0,1/2) representations (the complexification of the QLA is isomorphic
to the product of 2 copies of $U_qsl(2,C)$ ) would give an operation on the hierarchy of
partner sets, mapping each $R_{k,p}$ to $R_{k+1/2,p} \oplus
R_{k-1/2,p}$. Since tensor product is functorial, the morphisms of the
category, which we are using to construct our diagrams, are also acted
on by the supersymmetry functors.

This is the motivation for the name ``hypergravity''. If we think of
the labelling category for the TSS as an extension of the labels for
the model for GR, it has the appearance of a hierarchy of copies of
the GR labels, alternatively fermionic and bosonic, together with two
natural chiral fermionic maps relating adjacent elements of the hierarchy.
This is reminiscent of supersymmetric field theories, except that the
hierarchy seems to be infinite. The coupling of each ``hyperpartner''
to itself mirrors the gravity multiplet.

The above cited result about coupling of different partner sets is one
motivation for conjecture 1 above. For small curvatures, it would be
improbable to see k fluctuate by an entire half unit. Another piece of
evidence is the fact that if we restrict the models of [BC1,2] to
configurations corresponding to flat metrics, they in fact become
topological themselves [CKY1]. Of course the idea that ``nearly flat'' metrics
are ``nearly topological''needs a careful quantitative study.

At this point, I hope it is clear that ``what
would be the low energy effective field content of the hypergravity
multiplet?'' is a plausible question. One could try to find an analog
of the renormalization group for categorical state sums, and try to
discover how much of the hypergravity multiplet was relevant.

In the next section we explore
the possibility that, because of an algebraic coincidence of
noncommutative geometry which does not appear to have a classical
geometric analogue, the standard model might well emerge naturally
within hypergravity.

\bigskip

{\bf 3. Connections with the Standard Model and the Connes-Lott Model}

\bigskip

In [C-C], it was argued that the standard model emerges naturally
provided an algebra which the authors called the ``world algebra''
appeared as a symmetry algebra. The algebra they meant was 

\bigskip

$W= C \oplus H \oplus M^3(C)$ .

\bigskip

It was incidentally mentioned in their work, and later elaborated in
[Coq], that this algebra emerges as the semisimple part of the quantum
group $U_qsl(2,C)$ when we let q be a third root of unity.

Note that this is different from the common truncation at a 4rth root
of unity. This will be still another new algebraic question to study
in order to investigate this model.

Now let us remind ourselves that the QLA is a ``real form'' of $U_q
sl(2,C) \otimes U_qsl(2,C)$, more precisely, that its complexification
is exactly that algebra [BR1]. 

Hence a truncation of the QLA at a third root of unity would contain
two chiral copies of the ``world algebra''as its semisimple part. 

It is not unreasonable to expect that a nilpotent piece of an algebra
would vanish in a low energy (=long distance) limit, leaving the world
algebra as the effective symmetry of the theory.

This suggests that it might be interesting to examine other odd roots
of unity as well, to see if they predict hierarchies of fields which
include both the standard model and others which might be more
massive, but potentially discoverable. 

It is not possible at this point to demonstrate that the standard
model actually emerges from the hypergravity multiplet at q a third
root of unity. The truncated representation theory needs to be
studied, to determine how the representations of our category
decompose under the action of the world algebra. In addition, a form of the renormalization group needs to be developed for
CSS's, and many hard calculations need to be done. Nevertheless, in
QFT symmetry is a very strong principle, and seeing the symmetry of
the standard model emerge by accident in the hypergravity picture is
very surprizing at least.

\bigskip

{\bf 4. Categorical Feynmanology}

\bigskip

In our proposal, there is no true (continuum) QFT except TQFT. The labels that
appear as matter fields in a certain regime merge into a topological
sum in the continuum limit. Thus the effective theory of matter should
be a combinatorial, algebraic one, which could appear within a tensor
category.

Such a point of view already exists in QFT as practiced,
namely, that of Feynman.

It has long been one of the disturbing problems of theoretical QFT
that neither the derivation of the Feynman rules from a QFT nor the
reconstruction of a QFT from Feynman amplitudes is mathematically
sound. On the other hand, it is rather the Feynman rules than the
continuum QFT which have directly been verified by experiment.

All this makes us wonder if the Feynman diagrams of the physically
relevant QFTs could appear directly within the tensor categories we
are using in our constructions.

This proposal is not out of harmony with the original
ideas of Feynman [F], who seems to have believed that fundamental quantum
processes were discrete, and to have been rather sceptical of
continuum QFT as a framework. The point of view that perturbative QFT
simply sidestepped unknown physics that appears at a very high energy
has been a popular one for the subsequent development of perturbation theory.

We are proposing that a categorical state sum is the most natural
candidate for the unknown high energy Physics.

We want to present some lines of argument to make this plausible.

In the first place, the Feynman rules have a very close formal
similarity to the structure of a tensor category. Morphisms in a
tensor category can be constructed using graphs, with objects on edges
and tensor operators on vertices. Furthermore, duality in tensor
categories gives an identification of the set of vertices with a
certain number of lines in and the rest out with the vertices with the
same number and type of lines, but the separation into in and out
lines changed. For example, $Hom(A \otimes B, C \otimes D)=Hom (A
\otimes C^*,B^* \otimes D)$. This is formally identical to Feynman's
use of CPT invariance to identify vertices with edges going forward in
time with similar edges going backwards.

The similarity between Feynmanology and categorical diagrams is not
accidental. When one constructs a Lagrangian, one first determines
what fields one needs to include, then arranges them as
representations of the symmetries the theory is expected to
have. Next, one looks for scalar combinations of the fields to include
in the Lagrangian. These provide the possible vertices of the
Feynmanology of the theory.

Note that representations of a group or algebra form a tensor
category, and that by duality, scalars in the theory correspond to
tensor operators. Thus, up to this point, the construction of tensor
categories and Feynmanologies is identical.

Then there is the question of the Feynman integral, which gives us an
evaluation of a Feynman diagram. In [B, F-K], it was pointed out that
the evaluations of the closed diagrams in the categories of
representations for Euclidean 4D GR (relativistic spin nets), were
exactly Feynman integrals. In [BC2, R-R1 ], the same was shown for a
regularized definition of an evaluation for closed diagrams in the
category of unitary representations of the Lorentz algebra. 

It might seem that the analogy between Feynmanology and tensor
categories breaks down when we evaluate an open Feynman graph and get
a number rather than a morphism. This is not correct. The evaluation
of a Feynman graph is a function of momenta for the incoming and
outgoing particles. Passing to the momentum representation for the
state spaces, we can regard these numbers as the matrix elements of a
linear map. It is natural to think of this map as a morphism in a
tensor category. The Ward or Slavnov-Taylor relations [t'H] amount to
the statement that this map is an intertwiner for the gauge algebra,
i.e. that it lives in the category of representations of the total
symmetry of the system.

In the standard construction of a QFT, we would now select a few
fields as the fundamental ones, find which vertices are
renormalisable, and adjust the coupling constants in front of them to
fit experiment. The picture of the vacuum in Feynmanological
QFT is then in effect a subsummation of a categorical state sum, in
which only special terms are counted, and with special numerical
weights.

We now want to make a specific conjecture as to how Feynmanology as we
know it arises.

 {\bf CONJECTURE OF CATEGORICAL FEYNMANOLOGY: Let us postulate that just below the scale at which
the hypergravity model goes topological, we have all the objects and morphisms of
the category of representations of the physical
symmetries of the standard model appearing in an effective state sum. This would correspond
to a sum over Feynman diagrams, but allowing all possible particles
and vertices. Let us then assume that the renormalization group acts on
this theory, making all the unobserved types of particles become
unstable in the low energy limit, and all the unrenormalizable
vertices irrelevant (i.e. shrinking their coupling constants to near 0)
and giving the remaining coupling constants their physical values.

The Feynman picture of the vacuum then appears as the low energy
surviving remnant of the full categorical sum. }

The remarkable thing about this conjecture is that it is really not so
terribly radical. The idea that ``all terms not forbidden by a
symmetry appear in the effective Lagrangian'' is commonplace in
QFT. Particles are considered to be completely specified by their
``quantum numbers'' which just specify the representation of the
physical symmetry which they form. The idea that the observed
interactions are determined by the renormalization group is also
standard.

It is not clear whether the conjecture above, which boils dow to the
assumption that all possible fields appear at the fundamental scale
and all vertices appear with equal coupling constants, has any
physically testable consequences. It is arguable that it removes an
esthetic problem from QFT. There has been the feeling that the Feynman
revolution of the late 40s was disturbing in that it consisted of
calculational tricks rather than a conceptual leap. If we really
should think of summing over the entire category of representations of
the total physical symmetry rather than merely the low energy limit, then Feynman in
effect discovered the relationship between combinatorial topology and
categorical algebra, which in some eyes is an elegant and fundamental
departure from continuum Physics after all.

We are led to the idea that the Feynmanology of the standard model
could appear directly as terms in our hypergravity TSS corresponding
to representations of the world algebra, which, if our conjectures are
true, would become important in an expansion around the constrained
model in a low energy regime.

This suggests that although the strong force is nonperturbative in the
infrared, physical calculations using perturbation theory for QCD could
actually be done, since the onset of the topological theory would cut
off the perturbation series. Terms with a large but finite number of
loops would dominate the physics. Summing them would be equivalent to
a categorical state sum on a triangulation with Planck scale spacings.

It is very tempting, in such a model, to take the ideas of Feynman more
literally. Perhaps we might think of modifying our quasi-vacuum to
include terms which fill the Dirac sea. It is also very natural to
think of inserting Feynman vertices, with their invariance wrt time
ordering into state sums representing general relativity, where
different terms would correspond to different causal orderings.

This picture suggests that we attempt to generalize Feynmanology to
the categorical setting. We could attempt to investigate such
fundamental ideas of particle physics as renormalizability, anomalies,
and current algebras in the greater generalization of a tensor
category for example, using representations of a quantum group. Such a study would be inherently interesting, and might also
help us to further the program of this paper.

\bigskip

{\bf 5. Stringy Vacua}

\bigskip

It was observed by Lee Smolin [S], that the natural perturbation
theory around a classical solution in the Euclidean signature
categorical state sums we are studying would resemble
propagation of a string. This is because the natural basic perturbation to
make to a labelled diagram representing a morphism in the category of
representations of  $U_qsu(2)$ is to tensor by a constant spin around
a loop. In fact, the binor calculus shows that any ``spin net'' as
such diagrams are also known, is a superposition of such minimal
perturbations.

It is not immediately obvious that this stringy picture would work as
well in the Lorentzian signature, but that is a common enough problem
in string theory.

Thus, if we want to study perturbation theory around a fixed term in a
categorical state sum, we insert loops with spin labels, and then ask
the probability that some other combination of basic perturbation
loops appears later. This can be calculated by tracing the different
possible paths for the loops, i.e. by summing over discrete world
sheets for a discretized string.

Assuming for the moment that such a perturbation theory has some range
of validity (a non-trivial assumption, but at least one which could be
investigated), in what sort of space would it appear? 

The CSSs we are using in this paper all begin by putting
representations on the 2-simplices of a triangulated 4-manifold. Thus,
the loops of representations we use as perturbations are occupying
little circles of 2-simplices. If we think of the simplices as small,
we can approximate these as circles in the bundle of bivectors over
the spacetime 4-manifold.

Now the bundle of bivectors over a 4-manifold is a 10 dimensional
manifold, fibered over the spacetime with a 6 dimensional fiber.

The natural suggestion is that in the limit of finer triangulations
the perturbation theory would look like 10 dimensional string
theory. Since the hypergravity multiplet contains both bosonic and
fermionic terms with a natural functor connecting them, it is tempting
to think it would reproduce superstring theory.

This further suggests that geometries in 10 dimensions could serve as
quasivacua for the theory. One would decide what representation to
assign to a 2-simplex by finding its area viewed as a surface in the
bivector space endowed with a suitable metric. 

We seem to be suggesting that the extra 6 dimensions for compactified
models should be thought of not merely as small, but as infinitesimal,
i.e. as bivectors associated to the tangent space.

This conjecture is principally motivated by the enormous work already
existing on superstrings and string vacua. The only connections that
are really solid at this point are the coincidence of dimensions, the
hierarchy of bosonic and fermionic excitations, and the 
``stringiness'' of the picture.

\bigskip

{\bf 6 Generalizations. Membrane theory?}

\bigskip

In broad outline, we are suggesting that the combination of a TSS and
some interesting constrained subtheory might provide a mathematical
setting for quantum physics. We have, in effect, begun to study the simplest
possible such combination, corresponding to the lowest rank of
noncompact quantum groups. It is intriguing that there is some
indication that the standard model might emerge from the simplest 
example.

However, there will certainly be many other similar models which one
could investigate. If the idea of obtaining the world algebra from the
hypergravity multiplet didnt work, one could try bringing a gauge
group directly into the theory by starting with a larger noncompact
quantum group, for example. If nothing else, this would pose some very
natural questions for representation theory.

Another possible generalization comes from the fact that the four
dimensional TSS of [CKY1] from which we started is not the most general
categorical construction possible. This is because it begins from a
braided tensor category, whereas the most general 4D construction
would use a spherical tensor 2-category [M1,2] or else a Hopf category [CF]. The
construction we are using is a special case, because a braided tensor
category is a 2-category with one object.

The most general construction puts objects from the 2-category on
edges of the triangulation, and 1-morphisms on faces. Since a braided
tensor category is a 2-category with one object, the labelling of the
edges becomes trivial and disappears.

In terms of the diagrammatic description, the edges of the diagram
become boundaries of surfaces, which carry new labels, in the most
general case.

At this point, I do not know of any physically significant tensor
2-categories. Interesting mathematical examples are constructed in
[M2].

However, it is amusing to ask what would happen if we tried to find a
geometric picture for a perturbation theory for a TSS model based on a
general spherical tensor 2-category. The edges would correspond to
bivectors as above, while the surfaces bounding them would live in the
space of trivectors containing a bivector. The total space of such configurations is a 12
dimensional space. Thus, something like a membrane theory in 12d would be the
natural choice for a geometric background. This is of course extremely
speculative, but perhaps worth mentioning.

\bigskip

{\bf 7. Conclusions. A Convergence of Ideas?}

\bigskip

The direction we are mapping out is based on four ideas:

\bigskip
1. The fundamental geometry of spacetime is discrete, not continuous,
hence should be constructed from a model on a triangulated manifold.

\bigskip

2.The model should be constructed from the structure of a tensor
category, or if you prefer from Feynman integrals attached to the
simplices.

\bigskip

3. The continuum limit should be fixed by the coming into play of a
hierarchy of new fields, corresponding to more parts of the tensor
category, which remove the ultraviolet problems by including an entire
tensor category and hence becoming topological.

\bigskip

4. The largeness, flatness, and apparent complexity of the universe as we
see it is due to an initial fluctuation, which caused it to take an
apparent form more complex than the fundamental one.

\bigskip

Aside from the categorical formulation,these ideas are not new in theoretical physics. In particular, the
picture of a multiplet of fields which comes into play only at high
energies and softens out short distance behavior is very similar to
supersymmetric theories.

What is new here is the application of unitary representations of
noncompact groups, and the proposal to use the
structure of the tensor category they form, and q-deformations of it directly to generate a
physical model.

The model we are proposing is quite close to the spin foam picture [R-R2,
Ba], which itself grew out of the loop variable program [R-S]. The
most important difference between the two is the passage from a three
dimensional to a four dimensional symmetry group, due to the adoption
of a Lagrangian approach.

To the (perhaps somewhat partial) eye of the author, one of the
remarkable features of the hypergravity model is the number of earlier
ideas of twentieth century Physics to which it seems to give a home.

In addition to the spin foam/loop variables picture mentioned above,
and to the possible connection with superstring theory, one could
mention the old idea of Einstein [St] that the fundamental theory
should be discrete and algebraic, Feynman's recasting of particle
Physics as a discrete theory (as we can say now, of evaluations of
words in a tensor category), Witten's idea of a ``topological phase'',
and Connes derivation of the standard model.

Another intriguing connection is with the ideas of Majid [Maj]. Since
the QLA is the quantum double of a quasitriangular Hopf algebra, it is
a very canonical example of the crossed bialgebras he studies. Its
role in the models in this paper is therefore reminiscent of the
suggestions he makes relating them to quantum gravity. I do not know
if his deeper ideas about duality will appear more directly in the
development of this program.

Perhaps, even though each of these connections separately needs
considerable work to make it more precise, the sheer number of natural
connections to other streams of thought is worth noting.

We have made a large conceptual shift here from the mainstream of
quantum Physics. We are replacing continuum geometry and fields by
combinatorial geometry and categories of representations. The
justification for any fundamental reformulation must be that it opens
new possibilities for investigation.

It is much too soon to say that this program will solve the problems
of theoretical physics. It is fair to say that it contains many
directions to explore without obvious analogs in the continuum
picture, and with suggestive connections to physical
phenomenology. At a minimum, it can hardly fail to suggest interesting new
questions in categorical algebra.

\bigskip

{\bf BIBLIOGRAPHY}

\bigskip

[B] J. W. Barrett, { The Classical Evaluation of Relativistic Spin
Networks}, {\em Adv. Theor. Math. Phys.} {\bf 2} (1998), 593-600.

\bigskip

[BC1] J. W. Barrett and L. Crane, {\em J. Math. Phys.} {\bf 39} (1998) 3296-3302.

\bigskip

[BC2] J. W. Barrett and L. Crane, A Lorentzian Signature Model for
Quantum General Relativity gr-qc 9904025, to appear CQG

\bigskip

[Ba] J. C. Baez, { Spin Foam Models}, {\em Class. Quant. Grav.} {\bf 15} (1998)
, 1827-1858.

\bigskip

[B-R1] E. Buffenoir and Ph. Roche, Harmonic Analysis on the Quantum
Lorentz
group, q-alg 9710022

\bigskip

[B-R2] E. Buffenoir and Ph. Roche, Tensor products of Principal
Unitary Representations of the Quantum Lorentz Group and askey Wilson
polynomials math.QA 9910147

\bigskip

[CF] L. Crane and I. Frenkel, { Four Dimensional Topological Quantum Field
theory, Hopf Categories and the canonical Bases}, 
{\em J. Math. Phys.} {\bf 35} (1994), 5136-5154.

\bigskip

[C-C] A. Connes and A. Chamseddine The Spectral Action Principle hepth
9606001

\bigskip

[Coq] R. Coquereaux, On The Finite Dimensional Quantum Group
$M_3+M_{2|1}(Lambda^2)$ Lett. Math. Phys. 42 (1997) 309-328

\bigskip

[CKY1] L. Crane, L. Kauffman, and D. Yetter, { State Sum Invariants of 4-
Manifolds}, {\em JKTR} {\bf 6}(2) (1997), 177-234

\bigskip

[CY2] L. Crane and D. Yetter, unpublished.

\bigskip

[F] R. Feynman, Theory of Fundamental Processes W. A. Benjamin, 1962

\bigskip

[F-K]  L. Freidel and K. Krasnov, Simple Spin Networks as Feynman
Graphs.
hep-th/9903192

\bigskip

[G-N] I.M. Gelfand and M. A. Naimark, Unitary representations of the
proper
lorentz group, Izv. Akad. Nauk. SSSR 11 411 (1947)

\bigskip

[M1] M. Mackaay, { Spherical 2-categories and 4-Manifold Invariants},
{\em Adv. Math.} {\bf 143} (1999), 288-348.

\bigskip

[M2] M. Mackaay, Finite Groups, Spherical 2-Categories and 4-Manifold
Invariants, math-QA 9903003, to appear Adv. Math.

\bigskip

[Maj] S. Majid, Foundations of Quantum Group Theory Cambridge
University Press 1995

\bigskip

[N]  M. A. Naimark, Decomposition of a tensor product of irreducible
representations of the proper lorentz group into irreducible
representations, Am. Math. Soc. Translations ser 2 v 36

\bigskip

[P]W. Pusz, Irreducible unitary representations of the quantum
Lorentz group, Comm. Math. Phys. 152 571--626 (1993)
\bigskip

[P-Wo] P. Podles and S. L. Woronowitz Quantum deformations of the
Lorentz Group Comm. math. phys 136 399-432, (1990)

\bigskip

[R-R1] M. P. Reisenberger and C. Rovelli, Sum Over Surfaces Form of
Loop Quantum Gravity, Phys Rev. D56 (1997) 3490-3508

\bigskip

[R-R2] M. P. Reisenberger and C. Rovelli, { Sum over Surfaces Form of Loop
Quantum Gravity}, {\em Phys. Rev.} {\bf D 56} (1997), 3490-3508.

\bigskip

[R-S] C, Rovelli and L. Smolin, Spin Networks and Quantum Gravity
Phys. Rev. D 52 (1995) 5743-5759

\bigskip

[S] L. Smolin, Strings as Perturbations of Evolving Spin Networks
hepth 9801022

\bigskip

[St] J. Stachel, Einstein and Quantum Mechanics in Conceptual Problems
in Quantum Gravity Birkhauser 1991

\bigskip

[t'H] G. t'Hooft and M. Veltman diagrammar in Under the spell of The
gauge Principle, World scientific, 1994

\bigskip

[W] E. WittenTopological Quantum Field Theory and The Jones
Polynomial, Comm. Math. Phys. 121 (1989) 357

\bigskip

\end{document}